\documentclass[12pt]{article}

\usepackage{latexsym,amssymb,latexsym,latexsym,amssymb,overcite,epsf}

\newtheorem{Theorem}{Theorem}
\newtheorem{Definition}[Theorem]{Definition}
\newtheorem{Lemma}[Theorem]{Lemma}
\newtheorem{Corollary}[Theorem]{Corollary} 

\newenvironment{Proof}{Proof:}{$\Box$}

\newcommand{\ket}[1]{|#1\rangle} 
\newcommand{\bra}[1]{\langle #1|}
\newcommand{\cH}{{\cal H}}
\newcommand{\R}{{\mathbbm R}}
\newcommand{\N}{{\mathbbm N}}
\newcommand{\C}{{\mathbbm C}}
\newcommand{\Z}{{\mathbbm Z}}

\usepackage{bbm,epsf}
\title{\Large \textbf{Minimally-disturbing Heisenberg-Weyl symmetric
measurements using hard-core collisions of Schr\"odinger particles}}

\author{Dominik Janzing and Thomas Decker\\ 
{\small Arbeitsgruppe Quantum Computing}\\  
{\small Institut f\"ur Algorithmen und Kognitive Systeme}\\
{\small Fakult\"at f\"ur Informatik,}
{\small Universit\"at Karlsruhe (TH)}\\
{\small Am Fasanengarten 5, D-76\,131 Karlsruhe, Germany}\\
{\small \{janzing, decker\}@ira.uka.de}}

\date{February 17, 2006}
\begin{document}
\maketitle

\begin{abstract}
In a previous paper we have presented a general scheme for the
implementation of symmetric generalized measurements (POVMs) on a
quantum computer.  This scheme is based on representation theory of
groups and methods to decompose matrices that intertwine two
representations. We extend this scheme in such a way that the
measurement is minimally disturbing, i.e., it changes the state vector
$\ket{\Psi}$ of the system to $\sqrt{\Pi} \ket{\Psi}$ where $\Pi$ is
the positive operator corresponding to the measured result.

Using this method, we construct quantum circuits for measurements with
Hei\-sen\-berg-Weyl symmetry. A continuous generalization leads to a
scheme for optimal simultaneous measurements of position and momentum of a
Schr\"odinger particle moving in one dimension such that the outcomes
satisfy $\Delta x \Delta p \geq \hbar$. 

The particle to be measured collides with two probe particles, one for
the position and the other for the momentum measurement. The position
and momentum resolution can be tuned by the entangled joint state of
the probe particles which is also generated by a collision with
hard-core potential.  The parameters of the POVM can then be
controlled by the initial widths of the wave functions of the probe
particles.  We point out some formal similarities and differences to
simultaneous measurements of quadrature amplitudes in quantum optics.
\end{abstract}

%
%
\section{Introduction}
The question of how to implement quantum measurements is an important
issue of quantum information theory.  Even though the standard model of
quantum computers uses only one-qubit measurements in the computational
basis at the end or during the computation \cite{NielsenChuang}, other
measurements are also relevant for quantum information 
for several reasons.

In quantum computing, models have been proposed where collective
measurements on more than one qubit are necessary~\cite{Leung}.  In
Ref.~\citen{Bacon} a quantum algorithm is described which uses even
more general measurements than the usual von Neumann measurements,
i.e., they are not described by a family of mutually orthogonal
projections but by a so-called positive operator-valued measure
(POVM).

In non-computing applications of quantum information theory, like
future nano\-science, it may, for instance, be useful to implement
approximative simultaneous measurements of observables which are
actually incompatible when measured accurately.  An important example
would be the position and the momentum of a Schr\"odinger particle.

The implementation of generalized measurements is not trivial since
this is also true for the smaller class of von-Neumann measurements
\cite{PSPACE}.  Even though it is known that it is in principle
possible to reduce every POVM measurement to a von Neumann measurement
on an extended quantum system it is little known so far about how to
realize the required transformation by physical processes, particularly
when the implementation should disturb the quantum state
in a minimal way. In
Refs.~\citen{Calsamiglia} and~\citen{CalsamigliaDiss} the class of
POVMs is described which can be measured using linear optics.  Some
special POVM measurements on low-dimensional spaces are described in
Refs.~\citen{Renes, Englert, Reha, Sasaki, Takeoka, Ahnert}.

In
Ref.~\citen{altesPapier} we have described a general design principle to
implement symmetric POVMs 
on a quantum register where universal
quantum computation capabilities are available.  
However, these implementations are not minimally-disturbing.

Here we describe an extension of the theory of
Ref.~\citen{altesPapier} such that the symmetric measurements disturb
the state in a minimal way and apply it to the Heisenberg-Weyl group
in finite dimensions.  Since the latter actually defines  
a {\it family} of
groups it is desirable to have an implementation which is efficient in
the sense that its running time scales polynomially in the logarithm
of the dimension, i.e., with 
the number of qubits. We show that this can
indeed be achieved for Heisenberg-Weyl groups with a power of two as
dimension.

We then adapt the implementation scheme to the continuous situation.  The
corresponding POVM provides a good example for a measurement where the
feature of minimal disturbance makes sense: Given 
the motivation to
measure position and momentum of a particle in order to monitor its
motion  one
would clearly try to avoid disturbance as far as possible.

The general idea of the paper is to show that the 
finite dimensional circuits provide a paradigm for
the continuous variable implementation.  
By replacing the finite dimensional gates with appropriate analogues, 
we obtain also a possible measurement scheme
even though it is not a priori clear how the required ``gates'' 
could be 
implemented physically. 
However, we show that a modification
of the gate sequence could in principle be realized
by three hard-core scattering processes. 
This kind of idealized scattering in not unphysical
since
hard-core potential can be a useful approximation in many real
collision processes.

We proceed as follows. In the next section we recapitulate the
definition of POVMs and recall a general scheme for the
implementation of general measurements by orthogonal measurements. 
In Sec.~3 we define the symmetry of POVMs and
present a method for designing measurement algorithms for 
symmetric POVMs. In
Sec.~4 we consider the implementation for two special classes of POVMs
to illustrate the latter.  Explicitly, we consider POVMs on
qubits with cyclic symmetry groups and POVMs on $d$-dimensional
quantum systems with Heisenberg-Weyl symmetry.  In Sec.~5 we convert
the implementation to quantum systems with Hilbert spaces of infinite
dimension and describe a potential realization by scattering
processes on an abstract level.  
In Sec.~6 we compare this scheme to a quantum optical
implementation of simultaneous measurements for the quadrature
amplitudes.

%
%
\section{Minimally-disturbing implementation by\\ von-Neumann measurements} 
In this section we briefly outline a general scheme
\cite{NielsenChuang} for the minimally-disturbing implementation of a
POVM.
Consider a quantum system with Hilbert space ${\mathbbm C}^d$. A POVM
consists of $n$ operators $\Pi_j \in {\mathbbm C}^{d \times d}$ with
$\Pi_j \geq 0$ and $\sum_j \Pi_j = I_d$ where $I_d$ denotes the
identity matrix of size $d \times d$. A definition for POVMs on
infinite dimensional quantum systems and an infinite number of results
can be found in Ref.~\citen{Davies}. In Sec.~5 we use this more
general definition but here we start with finite POVMs
since we consider implementation schemes on {\it quantum computers} 
at first.
Following Refs.~\citen{Barnum} and~\citen{BarnumDiss} we define:
\begin{Definition}[Minimally-disturbing measurement]
Let $(\Pi_j)$ be a POVM. Then a measurement is called minimally-disturbing
if it changes the state vector according to
\[
\ket{\Psi} \mapsto \frac{\sqrt{\Pi_j} \ket{\Psi}}
{\| \sqrt{\Pi_j} \ket{\Psi} \|}\,,
\]
given that the measurement result is $j$. 
\end{Definition}
The motivation for this definition is given by 
a theorem in Ref.~\citen{Barnum} stating that
the above type of measurements maximizes the average fidelity between
the input and the output state if the input is drawn from a uniformly
distributed ensemble of pure states.

The following lemma~\cite{NielsenChuang} 
reduces the implementation of this kind of measurements
to von-Neumann measurements (with L\"{u}der's projection 
postulate) in the standard basis.

\begin{Lemma}[Reduction of a POVM to a von-Neumann measurement]
\label{lem red}
Let $P$ be a POVM with the $n$ operators $\Pi_j \in 
{\mathbbm C}^{d \times d}$. Furthermore, let ${\mathbbm C}^n$ be
the Hilbert space of an ancilla that is initialized with
$\ket{0}$. Then, a minimally-disturbing measurement of $P$ can
be achieved by a measurement in the standard basis of the
ancilla after the implementation of a unitary 
$U \in {\mathbbm C}^{dn \times dn}$ satisfying
the equation
\begin{equation}\label{Eq 1}
U (\ket{0} \otimes \ket{\Psi}) = \sum_{j=0}^{n-1}
\ket{j} \otimes \sqrt{\Pi_j} \ket{\Psi}.
\end{equation}
\end{Lemma}
Eq.~(\ref{Eq 1}) states that $U$ is a unitary extension of the matrix
\begin{equation}\label{matrix m}
M:=\sum_j \ket{j}\otimes \sqrt{\Pi_j} = 
(\sqrt{\Pi_1}, \sqrt{\Pi_2},\dots,\sqrt{\Pi_n})^T \in 
{\mathbbm C}^{dn \times d}
\end{equation}
which is defined by $P$. In the following section we consider this
extension for symmetric POVMs. In some cases, the following
observations help to show that a unitary $U$ implements a POVM in a
minimally disturbing way.
\begin{Lemma}[Linear assignment of Kraus operators]
\label{Lemma kraus}
Let $U\in{\mathbbm C}^{dn \times dn}$ be a unitary operating on
a bipartite system which is initialized with $\ket{\Phi} \otimes
\ket{\Psi}$ where $\ket{\Phi} \in {\mathbbm C}^n$ and
$\ket{\Psi} \in {\mathbbm C}^d$.
Let the first component be measured in the standard basis
after the joint system has been subjected to the unitary $U$. 
Then the conditional post-measurement state is pure. 

Let $A_{U,\Phi,j}$ be the Kraus operator describing  
the corresponding state change 
\[
\ket{\Psi} \mapsto \frac{A_{U,\Phi,j} \ket{\Psi}}
{||A_{U,\Phi,j} \ket{\Psi}||}
\]
where $j$ is the measurement result. 
For each  $U$ and $j$, the mapping $\ket{\Phi} \mapsto
A_{U,\Phi,j}$ is linear.
\end{Lemma}

The proof is straightforward since the projected state
of the composed system is a product state and the map given by the
partial trace is linear. We find:

\begin{Corollary}[Minimally disturbing Kraus operators]
\label{Corr}
Let $P$ be a POVM with operators $\Pi_j$.  Furthermore, let
$U$, $\ket{\Phi}$, and $A_{U,\Phi,j}$ be as defined in 
Lemma~\ref{Lemma kraus}.  If the equation
\[
A_{U,\Phi,j}=\sqrt{\Pi_j}
\]
holds for all $j$
then $U$ gives rise to a 
minimally-disturbing measurement of $P$. In other words, whenever 
$A_{U,\Phi,j}$ is positive for each $j$, it defines a minimally-disturbing
measurement for the POVM given by
\[
\Pi_j:=A_{U,\Phi,j}^2\,.
\]
\end{Corollary}

%
%
\section{Implementation of symmetric POVMs}\label{sec:symm}
In this section we analyze how the symmetry of a POVM can be used for
the implementation scheme of Lemma~\ref{lem red}. Here, we follow the
approach of Ref.~\citen{altesPapier} where we have obtained a general
implementation scheme for POVMs without consideration of the
disturbance of the measurement process. This implementation scheme
also relies on the unitary extension of a matrix that is defined by
the POVM operators. It turned out, that the symmetry of the POVM leads
to a symmetry of the matrix which can be exploited for the
extension. In this section we show that a similar construction is
possible for the minimally-disturbing implementation of POVMs.

To begin with, we define the symmetry of POVMs:
\begin{Definition}[Symmetric POVMs]
Let $\sigma:G \to {\mathbbm C}^{d\times d}$ be a unitary
representation of a finite group $G$. A POVM with operators
$\Pi_0, \ldots, \Pi_{n-1}$ is called $(\sigma,\pi)$-symmetric if there is a
permutation representation $\pi:G \to S_n$ of the indices such
that
\[
\sigma(g) \Pi_j \sigma(g)^\dagger = \Pi_{\pi(g)j}\,.
\]
Here, $S_n$ denotes the symmetric group consisting of all permutations
of $n$ objects.
\end{Definition}
As mentioned above, the symmetry of a matrix is a useful tool
for the implementation of POVMs. Here we define the
symmetry of a matrix as in Refs.~\citen{Egner, Egner2, Pueschel}. 
\begin{Definition}[Matrices with symmetry and intertwining spaces]
Let $G$ be a finite group and $\sigma: G \to {\mathbbm C}^{m \times m}$ 
as well as 
$\tau: G \to {\mathbbm C}^{n \times n}$ be unitary representations.  
A matrix $A \in {\mathbbm C}^{m \times n}$ is 
$(\sigma,\tau)$-symmetric if it satisfies
\[
\sigma(g) A = A \tau(g)
\]
for all $g \in G$.  We also write $\sigma M = M \tau$ for the
$(\sigma, \tau)$-symmetry. We call the set
\[
{\rm Int}(\sigma,\tau)=\{ A \in {\mathbbm C}^{m \times n} : 
\sigma A = A \tau\}
\]
of all such matrices the intertwining space of $\sigma$ and $\tau$.
\end{Definition}

The structure of the intertwining space of two representations can be
easily specified if both representations are decomposed into a direct
sum of irreducible representations of the group as the following
lemma~\cite{Pueschel} shows:
\begin{Lemma}[Structure of intertwining space]\label{lem int}
Let $\sigma$ and $\tau$ be decomposed into the direct sums
\[
\sigma=\bigoplus_j \left(I_{m_j} \otimes \kappa_j\right)
\quad {\it and} \quad
\tau=\bigoplus_j \left(I_{n_j} \otimes \kappa_j\right)
\]
of different irreducible representations $\kappa_j$ of the group $G$. Then
\[
{\rm Int}(\sigma, \tau)= \bigoplus_j \left({\mathbbm C}^{m_j\times n_j} 
\otimes I_{{\rm deg}(\kappa_j)}\right)
\]
where ${\rm deg}(\kappa_j)$ denotes the degree of $\kappa_j$. For
$m_j=0$ and $n_j=0$ we insert $n_j{\rm deg}(\kappa_j)$ zero columns or
$m_j {\rm deg}(\kappa_j)$ zero rows, respectively.
\end{Lemma}

The key observation used for the extension of the matrix $M$ from
Eq.~(\ref{matrix m}) to a unitary is that the symmetry of a POVM leads
to a matrix $M$ with symmetry.  This is summarized in the following
lemma which can be proved by direct calculation.
\begin{Lemma}[Symmetry of a POVM and its matrix]
If the POVM with operators $\Pi_1, \ldots, \Pi_n$ is
$(\sigma,\pi)$-symmetric then the corresponding matrix $M$ is
$(\sigma_\pi \otimes \sigma,\sigma)$-symmetric with the permutation
matrix representation $\sigma_\pi(g)= \sum_j \ket{\pi(g)j}\bra{j}$.
\end{Lemma}

The following theorem explicitly shows how the $(\sigma_\pi \otimes
\sigma, \sigma)$-symmetry of $M$ can be extended to a $(\sigma_\pi
\otimes \sigma, \sigma \oplus {\tilde B}^\dagger \sigma^\prime {\tilde
B})$-symmetry of $U$ where $\sigma^\prime $ is an appropriate
representation and ${\tilde B}$ a unitary.
\begin{Theorem}[Implementation of symmetric POVMs]\label{th erg}
Let $M$ be the matrix of Eq.~(\ref{matrix m}) for a
$(\sigma,\pi)$-symmetric POVM with symmetry group $G$. Let $A$ and $B$
be transformations that decompose $\sigma_\pi\otimes \sigma$ 
and $\sigma$ into
irreducible representations, respectively. Then there is a
representation $\sigma^\prime$ of $G$ such that $B\sigma B^\dagger\, 
\oplus \sigma'$
is equal to $A(\sigma_\pi \otimes \sigma)A^\dagger$ up to a
permutation of the irreducible components.  Furthermore, there is a
transformation $W \in {\rm Int}(A(\sigma_\pi \otimes \sigma)A^\dagger,
B\sigma B^\dagger \oplus \sigma^\prime )$ which is a unitary extension
of $AMB^\dagger$.  Then
\[
U:= A^\dagger W (B\oplus \tilde{B})  
\]
implements the POVM for every unitary $\tilde{B}$. The unitary $U$ is
$(\sigma_\pi \otimes \sigma, \sigma \oplus {\tilde B} \sigma^\prime
{\tilde B}^\dagger)$-symmetric.
\end{Theorem} 
\Proof{ We decompose $\sigma_\pi \otimes \sigma$ and $\sigma$ with the
unitaries $A\in{\mathbbm C}^{dn \times dn}$ and $B \in {\mathbbm C}^{d
\times d}$, i.e., we obtain the equations
\[
A(\sigma_\pi \otimes \sigma) A^\dagger = 
\bigoplus_j \left( I_{m_j} \otimes \kappa_j \right) \quad
{\rm and} \quad B\sigma B^\dagger = 
\bigoplus_j \left( I_{n_j} \otimes \kappa_j \right)\,.
\]
Therefore, the equation 
\[
\left(\bigoplus_j (I_{m_j} \otimes \kappa_j)\right) AMB^\dagger = 
AMB^\dagger \left(\bigoplus_j (I_{n_j} \otimes \kappa_j)\right)
\]
holds. Following Lemma~\ref{lem int} the matrix $N:=AMB^\dagger$ has
the decomposition
\[
N=\bigoplus_j \left(A_j \otimes I_{d_j}\right)
\]
with $A_j \in {\mathbbm C}^{m_j \times n_j}$ and $d_j:={\rm
deg}(\kappa_j)$.  From Th.~5 of Ref.~\citen{altesPapier} it follows
that $B\sigma B^\dagger$ can be extended to $A(\sigma_\pi \otimes
\sigma)A^\dagger$, i.e., the representations $A(\sigma_\pi \otimes
\sigma)A^\dagger$ and $B\sigma B^\dagger \oplus \sigma^\prime$ with
\[
\sigma^\prime = \bigoplus_j ( I_{m_j-n_j} \otimes \kappa_j)
\]
are equal up to a permutation of the irreducible components.  We
choose a unitary extension $W \in {\rm Int}( A ( \sigma_\pi \otimes
\sigma) A^\dagger, B \sigma {\tilde B} \oplus \sigma^\prime)$ of $N$.
This extension can be achieved by appending appropriate columns to the
right side of $N$ since the matrix $N$ has orthogonal columns.  We can
write $W=(N|{\tilde N})$ for this extension if we denote the new
columns by ${\tilde N}$. With this matrix we obtain for an arbitrary
unitary ${\tilde B}\in {\mathbbm C}^{n(d-1)\times n(d-1)}$ the unitary
extension
\[
A^\dagger(N| {\tilde N})(B \oplus {\tilde B}) =
(M|A^\dagger{\tilde N}{\tilde B})
\]
of $M$. $\Box$
}

As shown in the following section, the unitary $W$ of Th.~\ref{th erg}
can be chosen to be sparse for some POVMs.  Within the standard model
of quantum computing, this can be used for obtaining efficient
decompositions into elementary gates for the cases discussed in the
next section. Furthermore, there are methods known to decompose the
transformations $A$ and $B\oplus {\tilde B}$ into products of simpler
matrices \cite{Pueschel,Egner,Egner2}.

%
%
\section{Examples} 
In this section we explicitly construct quantum circuits 
and implementation schemes for
the minimally-disturbing implementation of two families of symmetric POVMs.
First, we introduce the following notations:  For $m\in \N$ define
$\omega_m:=\exp(-2\pi i/m)$ and let 
\[
X_m:=\sum_{j=0}^{m-1} \ket{ (j+1) \, {\rm mod} \, m}
\bra{j}\in {\mathbbm C}^{m \times m}
\]
be the cyclic shift of the basis vectors of an $m$-dimensional space.
Furthermore, define the diagonal phase matrix
\[
Z_m:=\sum_{j=0}^{m-1} \omega_m^j \ket{j}\bra{j}
\in {\mathbbm C}^{m \times m}
\]
and the Fourier transform
\[
F_m :=\sqrt{\frac{1}{m}} \sum_{j,k=0}^{m-1} \omega_m^{jk}\ket{j}
\bra{k} \in {\mathbbm C}^{m \times m}.
\]

We obtain the equalities $F_m X_m
F_m^\dagger = Z_m$ and $Z_m X_m=\omega_m X_m Z_m$ which we will use in
the following without proof.

\subsection{Cyclic groups}
Simple examples for our implementation scheme are POVMs operating on a
qubit with a cyclic symmetry. Measurements with cyclic symmetry
can, for instance, provide an estimation of time when applied to a
dynamical quantum system.  The reason for this is that the time
evolution of a quantum system with energy eigenvalues being rational
multiples of each other is periodic and the dynamics is therefore a
unitary representation of $\rm{SO}(2)$.  This leads naturally to the
finite cyclic groups after discretization.

Fix $n \geq 2$. We consider the cyclic group $C_n
=\langle r: r^n =1\rangle$ with $n$ elements, the unitary matrix
representation $\sigma:C_n \to {\mathbbm C}^{2\times 2}$ with
$\sigma(j)=R_n^j$ for 
\begin{equation}\label{Mat rn}
R_n:=\pmatrix{1&0\cr 0&\omega_n} \in {\mathbbm C}^{2
\times 2},
\end{equation}
and the orbit of the vector $\sqrt{1/n}(1,1)^T \in
{\mathbbm C}^2$ with respect to this representation of $C_n$.  We
have the POVM operators 
\begin{equation}\label{PiDef}
\Pi_j := \frac{1}{n}\pmatrix{1&0\cr
0&\omega_n^j}\pmatrix{1&1\cr 1&1}\pmatrix{1&0\cr 0&\omega_n^{-j}} = 
\frac{1}{n}\pmatrix{1&\omega_n^{-j}\cr\omega_n^j&1}
\end{equation}
for
$j \in \{0, \ldots , n-1\}$. 

Applying the methods discussed in Sec.~\ref{sec:symm} we obtain
the following unitary for the minimally-disturbing implementation
of the POVM with cyclic symmetry.
\begin{Theorem}[Implementation of POVMs with cyclic symmetry]\label{cyc povm}
The POVM with the operators of Eq.~(\ref{PiDef}) can be implemented by the
unitary
\[
U := 
(F_n^\dagger \otimes I_2)X_{2n}^\dagger 
(I_n \otimes F_2)K^\dagger \in {\mathbbm C}^{2n \times 2n}
\]
where $K$ denotes the permutation matrix which is defined by
\[
K\ket{2j} = \ket{j} \quad {\it and} \quad K\ket{2j+1} = \ket{n+j}\,.
\]
\end{Theorem}

The rather technical proof can be found in the appendix.  The idea is
the following.  The transformation $A$ must diagonalize the cyclic
shift $X_n \otimes I_2$.  This can be achieved by $F_n^\dagger \otimes
I_2$.  Furthermore, we need the cyclic shift $X_{2n}^\dagger$ to
obtain the correct order of the irreducible representations. The
transformation $B$ is trivial since $\sigma$ is already diagonal.  The
remaining transformation $(I_n \otimes F_2) K^\dagger$ is the the
sparse matrix in the intertwining space.

If $n$  is a power of $2$ the ancilla system can be  a qubit register
and 
the unitary of Th.~\ref{cyc povm}
can be implemented efficiently as the following corollary states.
\begin{Corollary}[Circuits for cyclic POVM]
For $n=2^m$, $m\geq 1$, the unitary $U$ of Th.~\ref{cyc povm} can
be implemented efficiently with the circuit of Fig.~\ref{circ-cyc}.
\end{Corollary}
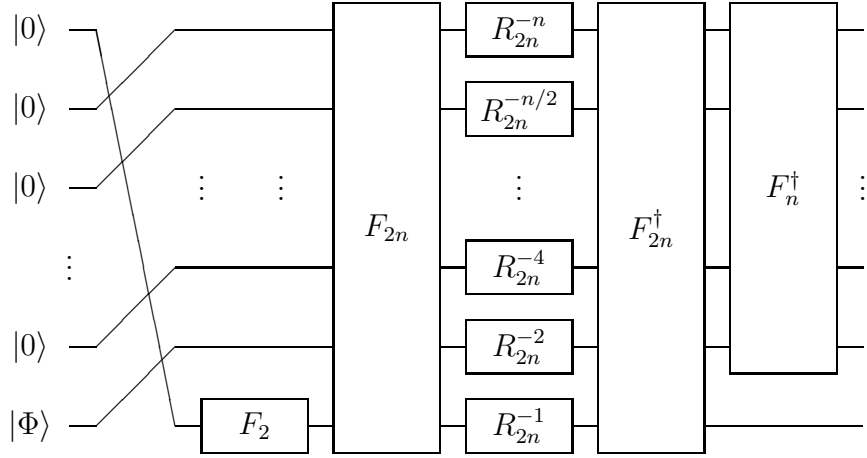
\begin{figure}
  \begin{center}
{
\unitlength=1.000000pt
\begin{picture}(315.00,170.00)(0.00,0.00)
\put(0.00,10.00){\makebox(0.00,0.00){$\ket{\Phi}$}}
\put(0.00,40.00){\makebox(0.00,0.00){$\ket{0}$}}
\put(0.00,100.00){\makebox(0.00,0.00){$\ket{0}$}}
\put(0.00,130.00){\makebox(0.00,0.00){$\ket{0}$}}
\put(0.00,160.00){\makebox(0.00,0.00){$\ket{0}$}}
\put(315.00,103.00){\makebox(0.00,0.00){$\vdots$}}
\put(185.00,103.00){\makebox(0.00,0.00){$\vdots$}}
\put(95.00,103.00){\makebox(0.00,0.00){$\vdots$}}
\put(65.00,103.00){\makebox(0.00,0.00){$\vdots$}}
\put(15.00,73.00){\makebox(0.00,0.00){$\vdots$}}
\put(105.00,10.00){\line(1,0){10.00}}
\put(305.00,160.00){\line(1,0){10.00}}
\put(305.00,130.00){\line(1,0){10.00}}
\put(305.00,70.00){\line(1,0){10.00}}
\put(315.00,40.00){\line(-1,0){10.00}}
\put(255.00,10.00){\line(1,0){60.00}}
\put(265.00,40.00){\line(-1,0){10.00}}
\put(255.00,70.00){\line(1,0){10.00}}
\put(265.00,130.00){\line(-1,0){10.00}}
\put(255.00,160.00){\line(1,0){10.00}}
\put(205.00,160.00){\line(1,0){10.00}}
\put(215.00,130.00){\line(-1,0){10.00}}
\put(205.00,70.00){\line(1,0){10.00}}
\put(215.00,40.00){\line(-1,0){10.00}}
\put(205.00,10.00){\line(1,0){10.00}}
\put(155.00,10.00){\line(1,0){10.00}}
\put(165.00,40.00){\line(-1,0){10.00}}
\put(155.00,70.00){\line(1,0){10.00}}
\put(165.00,130.00){\line(-1,0){10.00}}
\put(155.00,160.00){\line(1,0){10.00}}
\put(25.00,130.00){\line(-1,0){10.00}}
\put(55.00,160.00){\line(-1,-1){30.00}}
\put(25.00,100.00){\line(-1,0){10.00}}
\put(55.00,130.00){\line(-1,-1){30.00}}
\put(25.00,40.00){\line(-1,0){10.00}}
\put(55.00,70.00){\line(-1,-1){30.00}}
\put(25.00,10.00){\line(-1,0){10.00}}
\put(15.00,10.00){\line(0,1){0.00}}
\put(55.00,40.00){\line(-1,-1){30.00}}
\put(25.00,160.00){\line(-1,0){10.00}}
\put(55.00,10.00){\line(-1,5){30.00}}
\put(115.00,160.00){\line(-1,0){60.00}}
\put(115.00,130.00){\line(-1,0){60.00}}
\put(115.00,70.00){\line(-1,0){60.00}}
\put(115.00,40.00){\line(-1,0){60.00}}
\put(65.00,10.00){\line(-1,0){10.00}}
\put(65.00,0.00){\framebox(40.00,20.00){$F_2$}}
\put(265.00,30.00){\framebox(40.00,140.00){$F_n^\dagger$}}
\put(215.00,0.00){\framebox(40.00,170.00){$F_{2n}^\dagger$}}
\put(115.00,0.00){\framebox(40.00,170.00){$F_{2n}$}}
\put(165.00,0.00){\framebox(40.00,20.00){$R_{2n}^{-1}$}}
\put(165.00,30.00){\framebox(40.00,20.00){$R_{2n}^{-2}$}}
\put(165.00,60.00){\framebox(40.00,20.00){$R_{2n}^{-4}$}}
\put(165.00,120.00){\framebox(40.00,20.00){$R_{2n}^{-n/2}$}}
\put(165.00,150.00){\framebox(40.00,20.00){$R_{2n}^{-n}$}}
\end{picture}}

    \caption{\label{circ-cyc}Circuit for the
      implementation of POVMs with cyclic symmetry group $C_{2^m}$ on a qubit.
      The circuit operates on $m+1$ qubits. On the right side the
      upper $m$ qubits are measured in the standard basis.}
  \end{center}  
\end{figure}
\begin{Proof}
  Since Fourier transforms can be implemented with a polynomial
  number~\cite{NielsenChuang,KitaevAbelian} of elementary gates, i.e., one and
  two-qubit gates, $F_n^\dagger \otimes I_2$ can be implemented
  efficiently.  Furthermore, the cyclic shift $X_{2n}^\dagger$ can be
  written as $X_{2n}^\dagger = F_{2n}^\dagger Z_{2n}^\dagger F_{2n}$
  with
  \[ 
  Z_{2n}^\dagger= R_{2n}^{-n} \otimes R_{2n}^{-n/2} \otimes \ldots
  \otimes R_{2n}^{-1}\,.
  \]
  The unitary $K$ is only a cyclic shift of qubits.
\end{Proof}

\subsection{Heisenberg-Weyl groups}\label{Subsec:He}
The operators $X_d$ and $Z_d$ are discrete analogues of translations
in position and momentum space. If $\C^d$ denotes the $d$ possible
positions of a particle on a cyclic chain, the eigenvectors of $Z_d$
can be interpreted as positions eigenstates and the eigenvectors of
$X_d$ as eigenvectors of crystal momentum~\cite{Ziman}. Like for
continuous quantum systems these observables are incompatible and it
can be desirable to have approximative simultaneous measurements such
that the result can be interpreted as a point in $2d$-dimensional
``phase space''. In Sec.~5 we discuss the continuous analogue.

The basis of simultaneous measurements of position and momentum are
POVMs with Heisenberg-Weyl symmetry. For all $d \geq 2$, the
Heisenberg-Weyl group is given by $G=\langle X_d, Z_d \rangle$ and has
order $d^3$.  For a positive operator $\mu$ with ${\rm tr}(\mu)=1/d$
we consider the POVM with the $d^2$ operators
\begin{equation}\label{HWPOVM}
Z_d^k X_d^j \mu X_d^{-j} Z_d^{-k} \,\,\,\hbox{ for }\,\,\,
 k,j = 0, \ldots , d-1 \,.
\end{equation}
The following theorem shows how to implement this type of POVMs.

\begin{figure}
  \begin{center}
{
\unitlength=1.000000pt
\begin{picture}(290.00,80.00)(0.00,0.00)
\put(210.00,70.00){\circle*{5.00}}
\put(160.00,40.00){\circle*{5.00}}
\put(60.00,70.00){\circle*{5.00}}
\put(15.00,10.00){\makebox(0.00,0.00){$\ket{\Phi}$}}
\put(15.00,40.00){\makebox(0.00,0.00){$\ket{{\overline \alpha}}$}}
\put(15.00,70.00){\makebox(0.00,0.00){$\ket{\alpha}$}}
\put(60.00,50.00){\line(0,1){20.00}}
\put(160.00,20.00){\line(0,1){20.00}}
\put(210.00,70.00){\line(0,-1){50.00}}
\put(190.00,10.00){\line(-1,0){10.00}}
\put(280.00,70.00){\line(1,0){10.00}}
\put(280.00,40.00){\line(1,0){10.00}}
\put(230.00,10.00){\line(1,0){60.00}}
\put(140.00,10.00){\line(-1,0){110.00}}
\put(240.00,40.00){\line(-1,0){160.00}}
\put(130.00,70.00){\line(1,0){110.00}}
\put(240.00,60.00){\framebox(40.00,20.00){$F_d^\dagger$}}
\put(240.00,30.00){\framebox(40.00,20.00){$F_d^\dagger$}}
\put(190.00,0.00){\framebox(40.00,20.00){$Z_d$}}
\put(140.00,0.00){\framebox(40.00,20.00){$X_d^\dagger$}}
\put(90.00,70.00){\line(-1,0){60.00}}
\put(40.00,40.00){\line(-1,0){10.00}}
\put(90.00,60.00){\framebox(40.00,20.00){$F_d^\dagger$}}
\put(40.00,30.00){\framebox(40.00,20.00){$X_d^\dagger$}}
\end{picture}}

    \caption{\label{circ-hw}
      Schematic circuit for the POVM with Heisenberg-Weyl
      symmetry. The basis state $\ket{j}$ of the control wire causes
      the implementation of the $j$th power of the controlled
      operation. On the right side of the circuit the two upper
      systems are measured in the standard basis.}
  \end{center}
\end{figure}
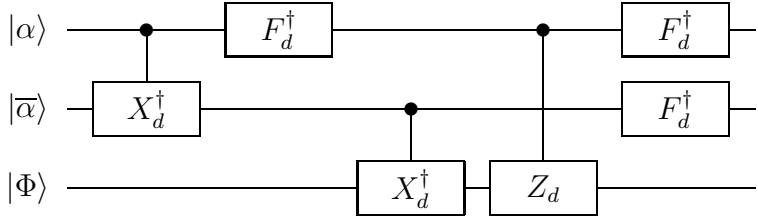

\begin{Theorem}[POVMs with Heisenberg-Weyl symmetry]\label{thm hw}
Given the Heisen\-berg-Weyl POVM $P$ with the $d^2$ operators from
Eq.~(\ref{HWPOVM}) with $\mu:=\ket{\alpha}\bra{\alpha}/d$ for some
state vector $\ket{\alpha}$.  Then $P$ can be implemented by the
circuit in Fig.~\ref{circ-hw} where the inputs of the ancillas are
given by $\ket{\alpha} \otimes \ket{\overline{\alpha}}$ with the
complex conjugated wave function ${\overline \alpha}$. For general $\mu$ the
ancilla input has to be replaced with the state vector
\[
\ket{\gamma}:=\sqrt{d} \sum_{j,k=0}^{d-1} \sqrt{\mu}_{jk} \ket{j}
\otimes \ket{k}\,\, \in {\mathbbm C}^{d^2}\,,
\]
where $\sqrt{\mu}_{jk}$ denotes the entry of $\sqrt{\mu}$ in
the $j$th row and $k$th column.
\end{Theorem}

The proof of the theorem can be found in the appendix. In the
following we briefly sketch the main points of the proof. For the
decomposition of $\sigma_\pi\otimes \sigma$ we observe that the
permutation $\pi$ given by the action of the Heisenberg-Weyl group on
the operators is a translation in the finite plane $(\Z/d \Z)^2$. This
translation is diagonalized by the inverse Fourier transform
$F_d^\dagger \otimes F_d^\dagger\otimes I_d$ at the end of the circuit
in Fig.~\ref{circ-hw}.  This transformation already block-diagonalizes
$\sigma_\pi\otimes \sigma$. However, the irreducible components are
only equivalent, but not equal, to $\sigma$. We apply the controlled
$Z$ and controlled $X^\dagger$ operations to obtain equality. Hence,
these operations correspond to the matrix $A$. The matrix $B$ is
trivial since $\sigma$ is an irreducible representation.  The unitary
extension $W$ used in Th.~\ref{thm hw} is decomposed into two
components. One component is given by the first two gates of the
circuit in Fig.~\ref{circ-hw}, the other is absorbed into the
preparation procedure for the initial state.

As already stated, we can efficiently implement $F_d$ by
elementary gates on a qubit register.  We also
obtain efficient implementations of controlled $X$ and $Z$ gates by
concatenations of controlled $R_d$ gates as defined in Eq.~(\ref{Mat
rn}).  Hence, we can implement the POVM with initial operator
$\mu=\ket{\alpha} \bra{\alpha}/d$ efficiently if the same is true for
the preparation of the states $\ket{\alpha}$ and $\ket{{\overline
\alpha}}$.

%
%
\section{Continuous measurements}\label{Sec 5}
Here we want to address how to implement
Heisenberg-Weyl symmetric POVMs 
for continuous quantum systems such that we have
also minimal disturbance. For a detailed mathematical 
description of such POVMs we
refer also to Refs.~\citen{Braunstein,Stenholm}.

The continuous degree of freedom can either be a Schr\"odinger wave of a quantum
particle moving on a line (where the Heisenberg-Weyl group formalizes
translations in position and momentum space) or a quantum optical
light mode (where the translations shift the quadrature amplitudes).
The most natural representation of the Hilbert space of a particle in
one dimension is $\cH:=L^2(\R)$, the space of square integrable
functions over the real line. For a light mode, it is often more
appropriate to choose the isomorphic Hilbert space $l^2(\N_0)$ of
square-summable sequences.  In this section, we will focus  on Schr\"odinger
particles since simultaneous measurements of quadrature amplitudes in
quantum optics have already been implemented \cite{Andersen}. We will
compare our implementation to the latter in the next section.

We  first describe the continuous analogues of the ``gates'' in
Fig.~\ref{circ-hw} and show that their concatenation leads 
indeed to a correct implementation.  Later we
will discuss a modification of the scheme which can be implemented by
hard-core scattering processes.  The description below refers to the
Schr\"odinger representation where the position operator $X$, defined
on a dense subspace of $\cH$, is the multiplication operator 
\[
X \psi (x) := x\psi (x) \,. 
\]
The momentum operator is
\[
P \psi (x) := -i \frac{d}{dx} \psi (x)\,, 
\]
where we have chosen the units such that $\hbar=1$.  Following
Sec.~3.4 of Ref.~\citen{Davies} (with a slight modification of the
sign) we introduce a family $(U_{s,t})$ of unitaries
\[
(U_{s,t}\psi) (x) := e^{-i xs} \psi(x-t) \,, 
\]
which formalize shifts in momentum and position space. 
These unitaries define a
measurement by the positive operators
\[
\Pi_{s,t}:=
\frac{1}{2\pi} U_{s,t}\ket{\alpha}\bra{\alpha} U_{s,t}^\dagger \,,
\]
where $\ket{\alpha} \in \cH$ is a wave function which is
sufficiently localized in momentum and position space.  The
probability density for the result $(s,t)$ is ${\rm tr}(\rho
\Pi_{s,t})$ if the system state is described by the density operator
$\rho$. The outcome $(s,t)$ is interpreted as momentum $s$ and
position $t$ of the particle in a ``coarse grained phase space''.  We
can clearly generalize the POVM above by replacing $ \ket{\alpha}
\bra{\alpha}/(2\pi)$ with any operator $\mu$ having trace $1/(2\pi)$.

In agreement with the discussions of finite POVMs in the preceding
sections we want to implement the POVM in such a way that the state
changes according to
\[
\rho \mapsto \frac{\sqrt{\Pi_{s,t}} \rho 
\sqrt{\Pi_{s,t}}}{\rm{tr}(\Pi_{s,t} \rho)}\,,
\] 
given that the measurement outcome is $(s,t)$.  

Now we describe how to find a continuous analogue of
the circuit in
Fig.~\ref{circ-hw}. The system Hilbert space 
$(\C^d)^{\otimes 3}$ is replaced by
$\cH^{\otimes 3}$, i.e., in additional to the
particle to be measured one uses two  particles in one dimension 
as ancilla system.  The final von Neumann measurement is a
position measurement on both ancillas\footnote{One could also use the
remaining two dimensions of a particle in three dimensions as ancilla
system.}.

The continuous analogues of the required gates are as follows.  The
discrete Fourier transform (whose inverse is occurring three times in
Fig.~\ref{circ-hw}) is replaced with the continuous unitary Fourier
transform
\begin{equation}\label{Four}
(F\psi) (x)= \sqrt{\frac{1}{2\pi}} \int_{-\infty}^{\infty}
e^{-i x y} \psi(y) dy\,.
\end{equation}
The controlled cyclic shift is replaced by a unitary $Y$ describing
controlled translations on the real line.  It acts on the wave
function $\psi$ of two particles according to
\begin{equation}\label{ConX}
(Y \psi)(x,y)=\psi(x,y-x)\,,
\end{equation}
since this transformation would correspond to the transformation
\[
\ket{x} \otimes \ket{y} \mapsto \ket{x} \otimes \ket{x +y}\,,
\]
if such position eigenstates $\ket{x}$ and $\ket{y}$ existed.
Conjugating $Y$ with the Fourier transform
on the second tensor component makes
it more apparent that it is indeed a unitary map since we obtain
then the multiplication operator
\begin{equation}\label{ConP}
V:=\Big( (I \otimes F) Y (I \otimes F^\dagger) \psi\Big) 
(x,y)= e^{-i xy} \psi(x,y)\,.
\end{equation}
Here $I$ denotes the identity operator on $\cH$.  The unitary in
Eq.~(\ref{ConP}) is the straightforward generalization of the
controlled phase-shift operation that is the fourth gate in
Fig.~\ref{circ-hw}. The following theorem shows that the above described
replacements provide in fact the desired measurement procedure:

\begin{Theorem}
\label{Infhw}
Replace the gates in Fig.~\ref{circ-hw} with their continuous
analogues as follows:
\begin{enumerate}

\item Set the inverse of the continuous unitary Fourier transform
given by Eq.~(\ref{Four}) instead of $F_d^\dagger$

\item Set the inverse of $Y$ given in Eq.~(\ref{ConX}) instead of the
controlled-$X_d^\dagger$ gate.

\item Set $V$ as given by Eq.~(\ref{ConP}) instead of the controlled-$Z_d$ gate.

\end{enumerate}

Let $\mu$ be an arbitrary positive operator with $tr(\mu)=1/(2\pi)$
and the two ancilla systems be in the state $\gamma$ with
\begin{equation}\label{def:gamma}
\ket{\gamma} := \sum_{j=0}^\infty 
\sqrt{\lambda_j} \ket{\alpha_j} \otimes \ket{\overline{\alpha}_j}\,,
\end{equation}
where $\ket{\alpha_j}$ is an eigenvector basis of $\mu$ such
that 
\[
2\pi \mu=\sum_j \lambda_j \ket{\alpha_j} \bra{ \alpha_j}\,.
\]
Then the resulting transformation on $\cH^{\otimes 3}$ implements a
minimally-disturbing measurement for the POVM
\[
\Pi_{s,t}:=U_{s,t} \,\mu \,U_{s,t}^\dagger \,,
\]
 when followed by position measurements on both ancillas
at the end and interpreting the position of the first particle
in Fig.~\ref{circ-hw} as $t$ and the position of the second as $s$.   
\end{Theorem}

\Proof{ Due to Lemma~\ref{Lemma kraus} and its corollary it is
sufficient to restrict the attention to rank-one operators
$\mu:=\ket{\alpha} \bra{\alpha}/(2\pi)$ and show that the unnormalized output
state coincides with the desired state.  The linearity argument holds
also if $\mu$ is an infinite series since one can check that the map
\[
\Phi \mapsto A_{U,\Phi,j}
\]
is continuous with respect to the topologies induced by the Hilbert space
norm and the operator norm, respectively. 
This is seen from
\[
\|\Phi\|^2={\rm tr}(A^\dagger_{U,\Phi, j}  A_{U,\Phi, j}) \leq 
\|A^\dagger_{U,\phi,j} A_{U,\Phi,j}\|=\|A_{U,\Phi,j}\|^2\,.
\]

The whole ``circuit'' creates
some wave function 
$\tilde{\Psi} \in \cH^{\otimes 3}$.
After measuring  $t$ and $s$ 
we obtain
an unnormalized conditional state vector
given by the wave function
\[
z \mapsto  \tilde{\Psi}(t,s,z):=\tilde{\Psi}_{t,s}(z)\,.
\]
We want to show that it satisfies
\[
\ket{\tilde{\Psi}_{t,s}}=\sqrt{\Pi_{s,t}} \ket{\Psi} = 
\sqrt{\frac{1}{2\pi}}   U_{s,t} 
\ket{\alpha} \langle \alpha |  U_{s,t}^\dagger \Psi \rangle \,.
\]
This means explicitly that 
\begin{equation}\label{Coin}
\tilde{\Psi}_{t,s}(z)= \sqrt{\frac{1}{2\pi}} 
 e^{-izs} \alpha (z-t) 
\int_{-\infty}^\infty\overline{\alpha} (u-t) e^{ius} \Psi(u) du\,.
\end{equation}

Now we calculate the effect of the circuit
starting with the joint state
\[
\alpha(x){\overline \alpha}(y)
\Psi(z)\,,
\] 
where $\Psi$ is the wave function of the measured particle.
First, we apply the controlled inverse translation and  obtain
\[
\alpha (x) \overline{\alpha} (y+x) \Psi( z)\,.
\]
The inverse Fourier transform changes this state to
\[
\sqrt{\frac{1}{2\pi}} 
\int_{-\infty}^\infty e^{i u x} \alpha (u) 
\overline{\alpha} (y+u) \Psi( z) 
du\,.
\]
The second controlled inverse shift followed by the controlled phase
yields
\[
\sqrt{\frac{1}{2\pi}} 
\int_{-\infty}^\infty e^{i u x} \alpha (u) 
\overline{\alpha} (y+u) e^{-ixz} 
\Psi( z+y) 
du\,.
\]
After applying  the inverse 
Fourier transform to both ancilla registers we obtain
\[
\sqrt{\frac{1}{8\pi^3}}\int_{-\infty}^\infty
\int_{-\infty}^\infty\int_{-\infty}^\infty e^{iy w} 
 e^{i xv} e^{i u v} \alpha (u) \overline{\alpha} (w+u) e^{-iv z} 
\Psi( z+w) 
du \, dv \, dw \,.
\]
We simplify this term into
\[
\sqrt{\frac{1}{8\pi^3}}
\int_{-\infty}^\infty\int_{-\infty}^\infty
\int_{-\infty}^\infty e^{iy w} 
 e^{i (x-z+u)v}  \alpha (u) \overline{\alpha} (w+u)  
\Psi( z+w) 
du \, dv \, dw \,.
\]
The integral over $v$ is only non-vanishing for $x-z+u=0$.  
Hence, we obtain
\[
\sqrt{\frac{1}{2\pi}} \int_{-\infty}^\infty e^{iy w} 
   \alpha (z-x) \overline{\alpha} (z+w-x)  
\Psi( z+w) 
dw \,.
\]
With the substitution $u:=z+w$ we get
\[
\sqrt{\frac{1}{2\pi}}   e^{-iyz} \alpha (z-x) 
 \int_{-\infty}^\infty e^{i y u} \overline{\alpha} (u-x)  
\Psi(u) 
du \,.
\]
The conditional state given that we obtain the result $x=t$ 
and $y=s$ coincides with Eq.~(\ref{Coin}).
}
$\Box$

\vspace{0.5cm}

\begin{figure}
  \begin{center}
{
\unitlength=1.000000pt
\begin{picture}(250.00,47.50)(0.00,0.00)
\put(220.00,40.00){\circle*{5.00}}
\put(220.00,40.00){\line(0,-1){20.00}}
\put(190.00,40.00){\line(1,0){60.00}}
\put(120.00,40.00){\circle*{5.00}}
\put(120.00,20.00){\line(0,1){20.00}}
\put(40.00,40.00){\line(1,0){110.00}}
\put(170.00,25.00){\makebox(0.00,0.00){$\cong :$}}
\put(240.00,10.00){\line(1,0){10.00}}
\put(200.00,10.00){\line(-1,0){10.00}}
\put(200.00,0.00){\framebox(40.00,20.00){$SC$}}
\put(-10.00,10.00){\makebox(0.00,0.00){light particle}}
\put(-10.00,40.00){\makebox(0.00,0.00){heavy particle}}
\put(140.00,10.00){\line(1,0){10.00}}
\put(90.00,10.00){\line(1,0){10.00}}
\put(40.00,10.00){\line(1,0){10.00}}
\put(100.00,0.00){\framebox(40.00,20.00){$X^2$}}
\put(50.00,0.00){\framebox(40.00,20.00){$R$}}
\end{picture}}

    \caption{\label{Fig:Corres}Correspondence between gates and a 
      scattering of two particles with extreme mass ratio. We shall call
the scattering a ``controlled $SC$ gate''.}
  \end{center}
\end{figure}
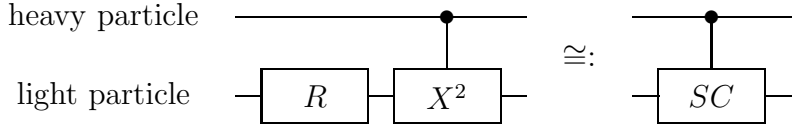

In order to realize the transformation in Th.~\ref{Infhw}
 by a physical process we first observe 
 that scattering processes realize quantum gates which are close
to the controlled phase shift in Eq.~(\ref{ConX}):
Consider two particles 
interacting with hard-core potential, i.e.,
the 
interaction energy is zero whenever their distance 
is larger than some $a>0$ and infinite if the distance is smaller than 
$a$. 
In Ref.~\citen{SchmueserScatt} we have discussed
the state change caused by such a scattering provided that
the considered time scale is small compared to the time scale
on which the width of wave packets grows by dispersion.  
We will first explain the scattering process in momentum space
since the change of momenta of classical particles provide
a good intuition about the quantum case. 
The momentum $p_2$ of the light particle obtains a sign change since it
is reflected.
Due to the conservation of total momentum,
the heavy particle acquires an additional momentum 
$2 p_2$. The vector of momenta of both particles 
is therefore changed according to a linear transformation $N$ 
given by
\[
N\left( \begin{array}{c} p_1 \\ p_2 \end{array} \right)
=\left( \begin{array}{c} p_1+2p_2 \\ -p_2 \end{array} \right)
\,.
\]
Neglecting irrelevant translations in position space,
the corresponding linear transformation $M$ in position space is already 
given by the requirement that the $4\times 4$ matrix 
transformation $M\oplus N$ acting
on the two positions and the two momenta has to be symplectic.
We have therefore $M=(N^T)^{-1}$ and 
obtain in agreement with Ref.~\citen{SchmueserScatt} 
\[
M=\left(\begin{array}{cc} 1 & 0 \\ 2 & -1 \end{array}\right)\,.
\]
The scattering process $S$ acts therefore on the wave function
in position space by multiplying the coordinate vector with $M$, i.e.,
\[
(S\psi)({\bf x}):= \psi (M {\bf x})\,, \,\,\,\,\,{\bf x} \in \R^2\,.
\]
We obtain 
\[
(S\psi)(x,y)=\psi(x,-y+2x)\,.
\]
In order to understand the relation to the gates in Th.~\ref{Infhw}, 
we may represent this operation by the circuit in Fig.~\ref{Fig:Corres}.
The ``reflection'' gate $R$ corresponds to a change of the wave function
according to 
\[
(R\psi)(x):=\psi(-x)\,.
\]

\begin{figure}
  \begin{center}
{
\unitlength=1.000000pt
\begin{picture}(260.00,80.00)(0.00,0.00)
\put(180.00,10.00){\circle*{5.00}}
\put(130.00,40.00){\circle*{5.00}}
\put(30.00,70.00){\circle*{5.00}}
\put(180.00,60.00){\line(0,-1){50.00}}
\put(30.00,50.00){\line(0,1){20.00}}
\put(130.00,20.00){\line(0,1){20.00}}
\put(150.00,10.00){\line(1,0){60.00}}
\put(0.00,10.00){\line(1,0){110.00}}
\put(50.00,40.00){\line(1,0){160.00}}
\put(200.00,70.00){\line(1,0){10.00}}
\put(100.00,70.00){\line(1,0){60.00}}
\put(60.00,70.00){\line(-1,0){60.00}}
\put(0.00,40.00){\line(1,0){10.00}}
\put(10.00,30.00){\framebox(40.00,20.00){$X^\dagger$}}
\put(265.00,40.00){\makebox(0.00,0.00){$P$ measurement}}
\put(265.00,70.00){\makebox(0.00,0.00){$X$ measurement}}
\put(160.00,60.00){\framebox(40.00,20.00){$X$}}
\put(110.00,0.00){\framebox(40.00,20.00){$X^\dagger$}}
\put(60.00,60.00){\framebox(40.00,20.00){$R$}}
\end{picture}}

    \caption{\label{Fig:Ansatz} Circuit equivalent to the circuit
      in Fig.~\ref{circ-hw}.}
  \end{center}
\end{figure}
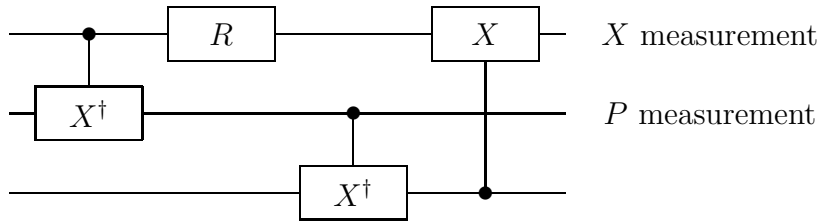

Elementary calculations show that Fig.~\ref{circ-hw}  is equivalent to
the circuit in Fig.~\ref{Fig:Ansatz} where we have absorbed the Fourier
transform on the  wire in the middle 
 by replacing an $X$-measurement with a $P$-measurement.

After we have converted the desired circuit into 
the equivalent one in  Fig.~\ref{Fig:Ansatz} that avoids controlled phase 
gates we still have the problem 
that it requires  controlled-$X$ gates 
and its inverse
instead of a 
controlled-$X^2$ gate which implements the shift twice. 
However, we observe that  
we may convert these gates into each other by conjugating them 
with 
the unitary squeezing operator 
\[
S_2\ket{x}:=\ket{2x}
\]
combined with reflections when needed.\footnote{Note that 
the described reduction of controlled-$SC$ to controlled-$X$ is also possible
in finite dimensions $d$. 
The definition $R|x\rangle:=|-x\rangle$ is always possible and
 $|x\rangle \mapsto |2x \;{\rm mod} \;d\rangle$
is  bijective if $d$ is odd. Then the ring
$\Z/d\Z$ allows division by $2$.}   
We will see later that we do not have to worry about the
physical realization of $S_2$ since we need this gate and its inverse 
 only
at the end or at the beginning of the first or the second wire. 
Hence, they can either be absorbed into the preparation procedure
or into the measurement by reinterpreting the result.

We will furthermore modify the entangling operation
on ancilla $1$ and $2$, i.e., the first gate 
of the circuit in Fig.~\ref{Fig:Ansatz} for the following reason.
An important feature of the circuits in Figs.~\ref{circ-hw} 
and~\ref{Fig:Ansatz} is that
the POVM consists of rank-one operators 
if the input ancilla state is the product state 
$\ket{\alpha}\otimes \ket{\overline{\alpha}}$ and 
entangled inputs lead to POVM operators 
of higher rank. The preparation of these entangled states
was not considered in Subsection~\ref{Subsec:He}.
Here we also want to describe {\it how
to entangle} ancilla $1$ and $2$ when POVMs of higher rank are desired. 
The goal is therefore to change the  operation 
on ancilla $1$ and $2$ preceding the interaction with
the system to be measured 
such that a family of {\it product} input states
allow the implementation of POVMs of higher rank.
In other word, we want to tune the achieved information and the caused 
disturbance of the measurements by plugging different product states 
into the circuit. 

\begin{figure}
  \begin{center}
{
\unitlength=1.000000pt
\begin{picture}(310.00,80.00)(0.00,0.00)
\put(0.00,10.00){\makebox(0.00,0.00){$m_3$}}
\put(0.00,40.00){\makebox(0.00,0.00){$m_2$}}
\put(0.00,70.00){\makebox(0.00,0.00){$m_1$}}
\put(30.00,10.00){\circle*{10.00}}
\put(30.00,40.00){\circle*{15.00}}
\put(30.00,70.00){\circle*{5.00}}
\put(250.00,10.00){\line(1,0){10.00}}
\put(210.00,0.00){\framebox(40.00,20.00){$R$}}
\put(180.00,10.00){\circle*{5.00}}
\put(130.00,40.00){\circle*{5.00}}
\put(80.00,40.00){\circle*{5.00}}
\put(80.00,60.00){\line(0,-1){20.00}}
\put(130.00,20.00){\line(0,1){20.00}}
\put(180.00,60.00){\line(0,-1){50.00}}
\put(110.00,10.00){\line(-1,0){60.00}}
\put(150.00,10.00){\line(1,0){60.00}}
\put(200.00,70.00){\line(1,0){60.00}}
\put(100.00,70.00){\line(1,0){60.00}}
\put(60.00,70.00){\line(-1,0){10.00}}
\put(260.00,40.00){\line(-1,0){210.00}}
\put(315.00,40.00){\makebox(0.00,0.00){$P$ measurement}}
\put(315.00,70.00){\makebox(0.00,0.00){$X$ measurement}}
\put(160.00,60.00){\framebox(40.00,20.00){$SC$}}
\put(110.00,0.00){\framebox(40.00,20.00){$SC$}}
\put(60.00,60.00){\framebox(40.00,20.00){$SC$}}
\end{picture}}

    \caption{\label{Fig:Finalcir} 
      Implementation of minimally-disturbing simultaneous measurement of 
      position and momentum by three scattering processes and one reflection. 
The masses $m_1$ and $m_2$  of the two ancilla particles are extremely small
or extremely large compared to the mass $m_3$ of the particle to be measured. 
}
  \end{center}
\end{figure}
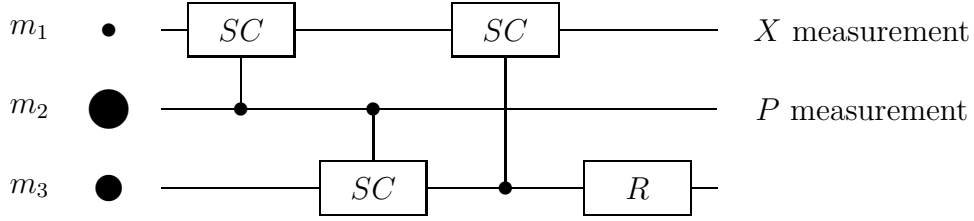

After subsequently replacing the gates in Fig.~\ref{Fig:Ansatz}
with scattering processes combined with squeezing operations 
and reflections
and modifying the entangling operation between the ancillas,
we found that an interesting class of POVMs can indeed
be implemented by three scattering processes
as depicted in Fig.~\ref{Fig:Finalcir}
when 
the initial ancilla states are Gaussian
wave packets.

To understand the effect of the ``circuit'' in Fig.~\ref{Fig:Finalcir}
we shall compute a $3\times 3$-matrix which describes
the effect of the whole circuit on the three position coordinates.
For doing so, we recall (see Fig.~\ref{Fig:Finalcir}) that  
the masses of the particles
satisfy
\[
m_1 \ll m_3 \ll m_2\,.
\]  
First, we implement a collision between particle $1$ and $2$. Here, 
the position of particle $2$ controls the position of particle
$1$. In analogy to the remarks above we describe the scattering and
reflection by matrices that acts 
on the vector of position coordinates
of the three particles.
The scattering processes with the pairs 
$(2,1)$, $(2,3)$, and $(3,1)$ correspond to the matrices  
\[
S_{21}:=\left(\begin{array}{ccc} 
-1 & 2 & 0 \\
0 & 1 & 0 \\
0 & 0 & 1 
\end{array}\right)\,, \hspace{1cm}
S_{23}:=\left(\begin{array}{ccc}
 1 & 0 & 0 \\
0 & 1 & 0 \\
0 & 2 & -1 
\end{array}\right)\,,
\hspace{1cm}
S_{31}:=\left(\begin{array}{ccc}
-1 & 0 & 2 \\
0 & 1 & 0 \\
0 & 0 & 1
\end{array}\right)\,.  
\] 
The scatterings are followed by a 
 reflection of the $z$-coordinate (the $R$ gate).
Taking into account that we have to concatenate the effect on the coordinates
from the left to the right,
the complete transformations in position coordinate space is given 
by
\[
A:=S_{31}\, S_{23}\, S_{21}\, R_3 =
\left(\begin{array}{ccc}
1 & 2 & 2 \\
0 & 1 & 0 \\
0 & 2 & 1 
\end{array}\right)\,.
\]
Let the initial state of the three particles be
given by the
wave function
\[
\alpha(x) \beta(y) \psi(z) \,.
\]
After subjecting the arguments to $A$ we obtain
\begin{equation}\label{Uphi}
\alpha(x+2y+2z)\beta(y) \psi(2y+z)\,.
\end{equation}
In order to reduce  a momentum measurement 
on the second wire to a position measurement we apply  a 
Fourier transform to the state (\ref{Uphi})
and obtain
\[
\sqrt{\frac{1}{2\pi}}\int_{-\infty}^\infty \alpha(x+2\tilde{w}+2z)\,
\beta(\tilde{w}) \,\psi(2\tilde{w}+z)\, e^{-i\tilde{w} y }
d\tilde{w}\,.
\]
With $w:=2\tilde{w}+z$ we get
\[
\sqrt{\frac{1}{8\pi}}\int_{-\infty}^\infty \alpha(x+w+z)\,\beta\Big((w-z)/2\Big) 
\,\psi(w)\, e^{i \frac{-y}{2}(w-z)}
dw\,.
\]
We define the integral kernel
\[
k_{x,y}(z,w):=\sqrt{\frac{1}{8\pi}}
\alpha(x+w+z) \,\beta\Big((w-z)/2\Big)\, e^{i \frac{-y}{2}(w-z)}\,.
\]  
It defines for fixed $x,y$ an operator $K_{x,y}$ 
on $\cH$ by
\[
(K_{x,y}\psi)(z) :=\int_{-\infty}^\infty k_{x,y}(z,w)\psi(w) dw\,.
\]
Note that the Kraus operators $K_{x,y}$  
describe 
the unnormalized output state $K_{x,y}|\psi\rangle$ of particle $3$ 
given that we have measured $x$ and $y$ on the first and second particle,
respectively (in straightforward analogy to the Kraus operators in  
Lemma~\ref{Lemma kraus} for the discrete setting). 

Now we show that $K_{x,y}$ can be obtained by subjecting $K_{0,0}$ to
the Heisenberg-Weyl group elements by
\[
K_{x,y}= U_{-x/2,-y} K_{0,0} U_{-x/2,-y}^\dagger\,.
\]
To see this, we observe that 
the translation by $-x/2$ in position space changes the integral  kernel
$k_{0,0}(z,w)$ into 
$k_{0,0} (z+x/2,w+x/2)$ and  
the additional translation  in momentum space by $-y/2$ 
changes it into
\[
k_{0,0} (z+x,w+x) e^{i\frac{-y}{2}(w-z)} =k_{x,y}(z,w)\,.
\]
This shows that the process in Fig.~\ref{Fig:Finalcir} implements
a measurement for the POVM
\[
\Pi_{s,t}:=U_{s,t} K_{0,0}^\dagger K_{0,0} U_{s,t} \,,
\]  
when reinterpreting the measurement outcomes $x,y$ on particle
$1$ and $2$ as $t=-x/2$ and $s=-y$, respectively. 
In order to obtain  a minimally-disturbing implementation, 
we have to ensure that $K_{0,0}$ is positive
(in straightforward generalization of Corollary~\ref{Corr} to
the continuous setting) because it can then be interpreted
as $\sqrt{\mu}$. 
If $\alpha$ and $\beta$ are real 
and $\beta$ is an even function, i.e., $\beta(-y)=\beta(y)$,
$K_{0,0}$ is self-adjoint due to
\[
k_{0,0}(z,w)=\overline{k_{0,0}(w,z)}\,.
\]
The integral kernel of $K_{0,0}$ is explicitly given by
\[
k_{0,0}(z,w)=\frac{1}{\sqrt{2\pi}}
 \alpha(w+z)\,\beta\Big((w-z)/2\Big)\,.
\]
Now we assume that $\alpha$ and $\beta$ are both real Gaussian
wave  functions
with widths $\sigma_1$ and $\sigma_2$, respectively,
i.e.,
\[
\alpha(x):=\frac{1}{\sqrt{\sigma_1}\pi^{1/4}}\,\exp\Big(
-\frac{x^2}{2\sigma_1^2}\Big)
\hspace{0.5cm}\hbox{ and }\hspace{0.5cm}
\beta(y):=\frac{1}{\sqrt{\sigma_1}\pi^{1/4}}\,\exp\Big(
-\frac{y^2}{2\sigma_2^2}\Big)\,.
\]
Under these conditions, $k_{0,0}$  
defines
a positive operator whenever $\sigma_1 \geq 2\sigma_2$. 
This follows from the following lemma after replacing $a$ and $b$ 
with $1/(2\sigma_1^2)$ and $1/(8\sigma_2^2)$, respectively.

\begin{Lemma}
The operator given by the integral kernel
\[
k(x,y):=d\, e^{-a(x+y)^2-b(x-y)^2}
\]
with $d>0$ 
is for all $b>a\geq0$  positive.
\end{Lemma}

\Proof{ Rewrite the kernel as
\begin{equation}\label{kDef}
k(x,y)=d\,e^{-2ax^2} e^{-(b-a)(x-y)^2}e^{-2ay^2}\,.
\end{equation}
It is known that the integral kernel
\[
\tilde{k}(x,y):=d\, e^{-c(x-y)^2}
\]
defines for all positive $c,d$ a positive operator \cite{Berg}
which we shall denote by $\tilde{K}$.
Then the operator $K$ given by the kernel (\ref{kDef})
can be written as $K=D\tilde{K}D$ where $D$ is the
multiplication operator
\[
(D\psi)(x):=e^{-2ax^2}\psi(x)\,.
\]
Hence, $K$ is also positive.
}
$\Box$

\vspace{0.5cm}

Since we have now described sufficient conditions for which
$K_{0,0}$ is positive, we would like to better understand 
the POVM operator $\mu=K_{0,0}^2$.
As simple computations show, it 
is (up to the normalization factor $2\pi$) 
given by
the reduced state of one particle in a two-particle system, if
the latter is described by the 
wave function
\begin{equation}\label{entG}
\phi(x,y):=\sqrt{2\pi}\,k_{0,0}=\alpha(x+y) \,\beta \Big((x-y)/2\Big)\,.
\end{equation}
It can be obtained 
from the state $\ket{\alpha}\otimes \ket{\beta}$ 
by a linear mapping of the wave function arguments
according to 
\[
\left(\begin{array}{c} x \\ y\end{array}\right)
\mapsto  
\left(\begin{array}{cc} 1 & 1 \\ 1/2 & -1/2 \end{array}\right)
\left(\begin{array}{c} x \\ y \end{array}\right)=:
G\left(\begin{array}{c} x \\ y \end{array}\right)\,.
\] 
Such a linear operation transforms the initial Gaussian state
into a entangled Gaussian state. Since Gaussian states
are completely determined by their covariance matrix \cite{Ferraro}
we will compute the latter for the state in Eq.~\ref{entG}.

For doing so,
we must describe the linear transformation corresponding to $G$ 
that acts on the arguments of the wave
 function in momentum space.
According to the  remarks at the beginning of this 
section, it is given by 
the transposed inverse: 
\[
\left(\begin{array}{c} p_x \\ p_y \end{array} \right)
\mapsto (G^T)^{-1} \left(\begin{array}{c} p_x \\ p_y \end{array} \right)\,.
\]
The covariance matrix of
a two-particle state $\rho$  consists of the entries
\[
{\rm tr}(\rho X_i X_j)-{\rm tr}(\rho X_i){\rm tr}(\rho X_j)\,\,\,\,\,
\hbox{ with }\,\,i,j=1,\dots,4\,
\]
where $X_1,X_2$ denote the position operators 
and $X_3,X_4$ the momentum operators of particle $1$ and $2$, 
respectively.
For the state
 $|\alpha\rangle \otimes |\beta\rangle$  
it is given by
(see 
Ref.~\citen{Ferraro})
\[
\sigma=
\left(\begin{array}{cccc}
\sigma_1^2/2 & 0 & 0 & 0\\
0 &  \sigma_2^2/2 &0 & 0 \\
0 & 0 &1/(2\sigma_1^2)  & 0\\
0 & 0 & 0 & 1/(2\sigma_2^2) 
\end{array} \right)\,.
\] 
If the coordinate vector in the position wave function is subjected to some
area-preserving linear map $G$ and the coordinates of the momentum wave
function to $(G^T)^{-1}$, the covariance matrix
transforms in the following way:
\begin{eqnarray*}
\sigma'&:=&\left(
\begin{array}{cc} G^{-1} & 0\\ 0 &G^T  \end{array}\right)
\sigma \left(
\begin{array}{cc} (G^T)^{-1} & 0\\0&  G  \end{array}\right)
\\
&=& 
\frac{1}{8}\left(\begin{array}{cccc}
\sigma_1^2 + 4\sigma_2^2 & \sigma_1^2-4\sigma_2^2 
& 0 & 0 \\
 \sigma_1^2-4\sigma_2^2 & \sigma_1^2 + 4\sigma_2^2& 
0 &
0  \\ 0 & 0 & \frac{4}{\sigma_1^2}+\frac{1}{\sigma_2^2} & 
\frac{4}{\sigma_1^2} -\frac{1}{\sigma_2^2}\\
0 & 0 & \frac{4}{\sigma_1^2} -\frac{1}{\sigma_2^2}&
\frac{4}{ \sigma_1^2}+\frac{1}{\sigma_2^2}
\end{array} \right)
\,,
\end{eqnarray*}
as simple computations show. 
The covariance matrix of the reduced state of  
each particle is given by the $2\times 2$ sub-matrices
that refer to its position and momentum. Due to the symmetry of our state,
it is for both particles
given by
\begin{equation}\label{Cov1}
\frac{1}{8}\left(\begin{array}{cc} \sigma_1^2 + 4\sigma_2^2 &0 \\
0 & 
 \frac{4}{\sigma_1^2} +\frac{1}{\sigma_2^2}
\end{array}\right)\,.
\end{equation}
It is known \cite{Ferraro} that such a state is pure if and only if 
the determinant is $1/4$. This is given for $\sigma_1=2\sigma_2$.
One can rewrite 
a Gaussian state 
of a single mode
having diagonal covariance matrix 
 as a thermal state of a harmonic oscillator\footnote{In 
quantum optics, one would also need squeezing transformations
to obtain a general diagonal Gaussian state. 
But here the product of frequency and mass
of the oscillator provides an additional free parameter.}
with frequency $\omega$, mass $m$ and average phonon number $N$.
It is explicitly given by
\[
\rho_{N, \nu} =(1-e^{-1/N})\sum_{n=0}^\infty e^{- n/N}  \ket{n}\bra{n}\,,
\]
where $\ket{n}$ with $n\in \N_0$ denotes the $n$th energy eigenstate
of the oscillator. 
We will first use 
dimensionless position and momentum variables
\begin{equation}\label{dimlessX}
X':=\frac{1}{\sqrt{2}}( a+a^\dagger)=\sqrt{m \omega} X
\end{equation}
and
\begin{equation}\label{dimlessP}
P':=\frac{1}{i\sqrt{2}}(a-a^\dagger)=\frac{1}{\sqrt{m \omega}} P\,,
\end{equation}
with creation operator $a^\dagger$ and annihilation operator $a$.
In these coordinates,  the covariance matrix of 
the thermal state with average phonon number $N$ is 
the identity matrix times $(N+1)/2$, 
this follows, e.g., from Eqs.~(2.16) in Ref.~\citen{Ferraro}. 
In natural units, we have therefore the covariance matrix 
\begin{equation}\label{Cov2}
\frac{N+1}{2} \left(\begin{array}{cc}
\frac{1}{m\omega} & 0\\ 0 & m\omega
\end{array}\right)\,.
\end{equation}
Comparing Eq.~(\ref{Cov1}) to Eq.~(\ref{Cov2}) we obtain
\[
(m\omega)^2=\frac{4/\sigma_1^2+1/\sigma_2^2}{\sigma_1^2+4\sigma_2^2}
\]
and
\[
(N+1)^2=\frac{1}{4}\Big(\sigma_1^2+4\sigma_2^2\Big)\Big(\frac{1}{\sigma_1^2}
+\frac{1}{4\sigma_2^2}\Big)\,.
\]
Hence
\[
N=\frac{1}{2} \sqrt{2+\frac{4\sigma_2^2}{\sigma_1^2}+
\frac{\sigma_1^2}{4\sigma_2^2}}-1\,.
\]
For $\sigma_1=2\sigma_2$ one obtains $N=0$, i.e., the ground state of the
oscillator which corresponds to a rank-one operator $\mu$. 
We rephrase the findings implied by the above discussion as a theorem:

\begin{Theorem}
Given three particles such that their masses satisfy
\[
m_1 \ll m_3 \ll m_2\,.
\]
Let the first and the second particle be in Gaussian states
with real wave functions such that  their 
widths satisfy $\sigma_1\geq 2 \sigma_2$.
Then the sequence of scattering processes depicted in 
Fig.~\ref{Fig:Finalcir} implements the Heisenberg-Weyl 
POVM  $(U_{s,t} \,\mu\, U_{s,t}^\dagger)_{s,t}$  
in a minimally-disturbing way when the position of particle
$1$ and the momentum of particle $2$ is measured 
and the result $(x,y)$ is interpreted as $t=-x/2$ 
and $s=-y$. The initial operator $\mu$ of the POVM
is given by 
\[
\mu=\frac{1}{2\pi}\rho_{N,m\omega}\,,
\] 
where
$\rho_{N,m\omega}$ is the thermal equilibrium state of
a harmonic oscillator with mass $m$ and frequency $\omega$
when the temperature is chosen such that the average phonon number is $N$.
The parameters $N$ and $m\omega$ are determined by the
widths $\sigma_1$ and $\sigma_2$ according to
\[
(m\omega)^2=\frac{4/\sigma_1^2+1/\sigma_2^2}{\sigma_1^2+4\sigma_2^2}\,,
\]
and
\[
N=\frac{1}{2} \sqrt{2+\frac{4\sigma_2^2}{\sigma_1^2}+
\frac{\sigma_1^2}{4\sigma_2^2}}-1\,.
\]
\end{Theorem}

We want to briefly explain qualitatively how the measured POVM is tuned by
the parameters $\sigma_1$ and $\sigma_2$. The ratio of both determine
the purity of $\mu$, for $\sigma_1=2\sigma_2$ we obtain a rank-one 
measurement. By increasing or decreasing both we can achieve
a better resolution in momentum space or in position space: 
Small values $\sigma_1,\sigma_2$  lead to good position measurements 
for the cost of having large errors in the momentum measurement.  
If $\sigma_1\gg 2 \sigma_2$ both position and momentum measurements
are bad and we obtain a measurement with small disturbance. 

For detailed discussions on the  disturbance and accuracy of the 
measurements we refer also to  Refs.~\citen{Braunstein,Stenholm}.
It is shown that the
outcomes for position and momentum
in POVMs of the above type satisfy 
the inequality $\Delta x \Delta p \geq
 \hbar$ in contrast to the Heisenberg uncertainty relation $\Delta x \Delta
 p \geq \hbar /2$.

%
%
\section{Comparison to quantum optics implementations}

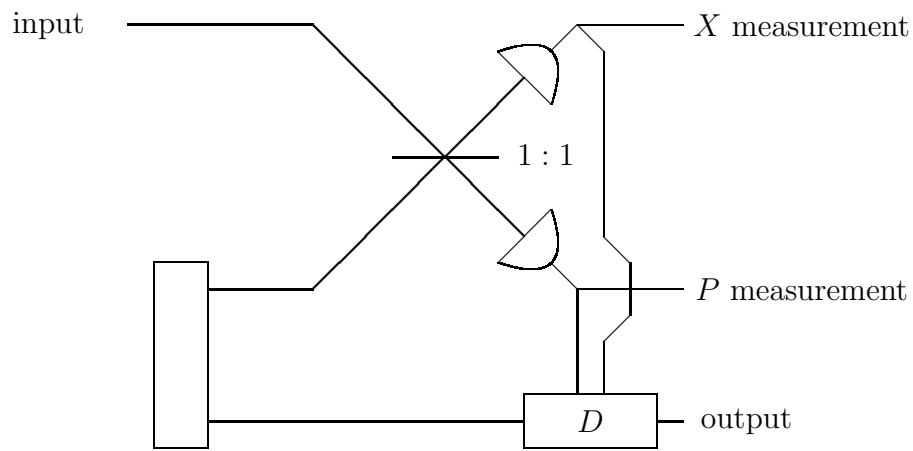
\begin{figure}
  \begin{center}
{
\unitlength=1.000000pt
\begin{picture}(285.00,160.00)(0.00,0.00)
\thicklines
\put(170.00,110.00){\line(-1,0){40.00}}
\put(230.00,10.00){\line(1,0){10.00}}
\put(60.00,10.00){\line(1,0){120.00}}
\put(100.00,60.00){\line(-1,0){40.00}}
\put(150.00,110.00){\line(-1,-1){50.00}}
\put(150.00,110.00){\line(1,-1){30.00}}
\put(180.00,140.00){\line(-1,-1){30.00}}
\put(100.00,160.00){\line(-1,0){70.00}}
\put(150.00,110.00){\line(-1,1){50.00}}
\thinlines
\put(210.00,150.00){\line(-1,1){10.00}}
\put(210.00,80.00){\line(0,1){70.00}}
\put(210.00,40.00){\line(0,-1){20.00}}
\put(220.00,50.00){\line(-1,-1){10.00}}
\put(220.00,70.00){\line(0,-1){20.00}}
\put(210.00,80.00){\line(1,-1){10.00}}
\put(188.00,111.00){\makebox(0.00,0.00){$1:1$}}
\put(0.00,160.00){\makebox(0.00,0.00){input}}
\put(264.00,10.00){\makebox(0.00,0.00){output}}
\put(285.00,60.00){\makebox(0.00,0.00){$P$ measurement}}
\put(285.00,160.00){\makebox(0.00,0.00){$X$ measurement}}
\put(40.00,0.00){\framebox(20.00,70.00){}}
\put(200.00,160.00){\line(1,0){40.00}}
\put(190.00,150.00){\line(1,1){10.00}}
\put(200.00,60.00){\line(1,0){40.00}}
\put(200.00,60.00){\line(0,-1){40.00}}
\put(190.00,70.00){\line(1,-1){10.00}}
\qbezier(170.00,70.00)(200.00,60.00)(190.00,90.00)
\qbezier(170.00,150.00)(200.00,160.00)(190.00,130.00)
\put(170.00,70.00){\line(1,1){20.00}}
\put(170.00,150.00){\line(1,-1){20.00}}
\put(180.00,0.00){\framebox(50.00,20.00){$D$}}
\end{picture}}

    \caption{\label{Fig:And}Measurement scheme of
      Ref.~\citen{Andersen}. The box at the beginning of mode $2$ and
      $3$ indicate the entangled input on these two modes. The
      entanglement tunes the POVM.  In the limit of infinite
      entanglement the output coincides with the input state and no
      information is gained. If mode $2$ and $3$ start in a product state,
      a rank-one POVM is implemented. The input interferes with mode
      $2$ in a balanced beam-splitter were one output mode is
      subjected to a position measurement and the other to a momentum
      measurement. The results determine the displacements
in position and momentum the output is
      subjected to.}
  \end{center}
  \label{circ andersen}
\end{figure}

There are meanwhile several methods known to measure
the quadrature amplitudes of a light mode simultaneously
(see Ref.~\citen{Andersen}). We will consider 
the scheme shown in Fig.~\ref{Fig:And}
 which has some nice  similarities to
the continuous analogue of our 
circuit in  Fig.~\ref{circ-hw}.
In the following, we will use the dimensionless
formal position and momentum operators as given in
Eqs.~(\ref{dimlessX}) and (\ref{dimlessP}) which 
generate the momentum and position translations in the
Heisenberg-Weyl group 
and define furthermore 
implicitly a Schr\"odinger representation of 
a single mode state as wave function $\psi \in L^2(\R)$ 
in position state. 

The method in Fig.~\ref{Fig:And} uses also two ancilla modes.
As in our proposal, the entanglement of the two modes tunes the 
POVM operator. 
One part of an entangled two-mode state 
(wire $2$ and $3$ in Fig.~\ref{Fig:And})
interferes with 
the input state (wire $1$) in a beam splitter. One of its output
modes is subjected to a position measurement, the other to a momentum
measurement. The results of the measurement determine furthermore
displacements performed on the second component of
the entangled input. The idea behind the scheme is to perform
a teleportation using a non-maximally entangled bipartite state
(a maximally entangled state 
does not exist anyway in continuous variables)
as resource. Then the transfer of quantum information 
is not perfect but the measurements 
performed during the ``bad'' teleportation provide
some information on the input state.  
Similar to our scheme,
the more entangled the joint state, the less information provides
the measurement and the less disturbance on the output state
will be observed. 

One technical difference to our scheme is that the input and output
are not on the same wire.   
The main difference is, however, that the 
interaction between input and entangled ancilla 
is given  by a beam-splitter, whereas we use scattering processes.
This is geometrically the difference between a rotation
or a shear in coordinate space (for details see Ref.~\citen{SchmueserScatt}).
Note, however, that the effect of the controlled displacements 
of Fig.~\ref{Fig:And} could be mimicked by  
a controlled-$X$ gate from wire $1$ to $3$ 
and a controlled-$Z$   
gate 
from wire $2$ to $3$ if 
the latter gate is conjugated by a Fourier transform on wire $2$.
The reason is that it does not make a difference whether 
the controlled operation is performed before
the measurement or afterwards. The scheme contains
therefore quite similar elements as ours. 

In Ref.~\citen{Andersen} we did not find an explicit remark saying
that their implementation scheme is minimally-disturbing in the sense
considered here since the authors use the term ``minimally-disturbing'' in 
a different sense. Furthermore, the attention was restricted to
Gaussian states for both the input as well as for the ancilla
states. We have observed that the implementation
is also applicable for non-Gaussian states and non-Gaussian POVM 
operators:

\begin{Theorem}
The scheme of Ref.~\ref{Fig:And} can in principle be used
for a minimally-disturbing implementation of any Heisenberg-Weyl symmetric
POVM 
\[
U_{s,t} \,\mu\, U_{s,t}^\dagger
\]
by preparing the ancilla state
\[
\sum_j \sqrt{\lambda}_j |\alpha_j \rangle \otimes |\overline{\alpha_j}\rangle \,,
\]
where the $|\alpha_j\rangle$ denote the eigenvector basis for $\mu$ 
and $\lambda_j/(2\pi)$   the corresponding eigenvalues.
\end{Theorem}

\Proof{
Due to the linearity argument in Lemma~\ref{Lemma kraus}  we may prove our
statement for the case that
the two ancillas are in a product state.
The initial three mode wave function is then given by
\[
\phi(x,y,z)=\psi(x)\,\alpha(y)\,\beta(z)\,,
\] 
where $\psi$ is the wave function of the mode to be measured.
The beam splitter transfers it to the wave function
\[
\psi\Big((x+y)/\sqrt{2}\Big)\,\alpha\Big((x-y)/\sqrt{2}\Big)\,\beta(z)\,.
\]
We simulate the momentum measurement by an inverse Fourier transform
followed by a position measurement. Conditioned on the 
measurement result $(x,y)$ we obtain therefore a
one-mode wave function  (having $z$ as argument) which is given by
\[
\sqrt{\frac{1}{2\pi}}\Big\{\int_{-\infty}^\infty 
\psi\Big((x+w)/\sqrt{2}\Big)\,\alpha\Big((-x+w)/\sqrt{2}\Big)\, 
e^{iwy}\,dw\Big\}\,\,\beta(z)\,.
\]
After obtaining the measurement results $x$ and $y$ on wire $1$ and $2$,
respectively,  
the conditioned displacement of position and momentum by 
$\sqrt{2}x$ and $\sqrt{2}y$, respectively, leads to
\begin{eqnarray*}
&&\sqrt{\frac{1}{2\pi}}\Big\{\int_{-\infty}^\infty 
\psi\Big((x+w)/\sqrt{2}\Big)\,\alpha\Big((-x+w)/\sqrt{2}\Big)\, 
e^{iwy}\,dw\Big\}\,\,
\beta(z-\sqrt{2}x)\, e^{-iz\sqrt{2}y}\\&&=
\sqrt{\frac{1}{2\pi}}\Big\{\int 
\psi(w)\,\alpha(-\sqrt{2}x+w)\, e^{iw\sqrt{2}y}dw \Big\}\,\,\,
\beta(z-\sqrt{2} x)\, e^{-iz\sqrt{2}y}
\,.
\end{eqnarray*}
This shows that the unnormalized state vector 
of the third 
particle, given that $x,y$ was measured, reads 
\[
U_{\sqrt{2}y,\sqrt{2} x}\ket{\beta} \bra{\overline{\alpha}}
U^\dagger_{\sqrt{2}y,\sqrt{2}x}\ket{\psi}\,.
\] 
After taking into account that the quantum optics convention for
position and momentum differs from the canonical definition
of Eq.~(\ref{dimlessX}) and (\ref{dimlessP}) by the factor $\sqrt{2}$
(see Ref.~\cite{Ferraro}), this is exactly the desired output state.

By choosing the input $\ket{\overline{\alpha}} \otimes 
\ket{\alpha}$ we
have therefore $\mu=\ket{\alpha} \bra{\alpha}=\sqrt{\mu}$.
Similarly we can obtain operators $\mu$ with higher rank by choosing
 entangled input states. 
}$\Box$

\vspace{0.5cm}
Note that the calculations which show that the scheme does indeed
implement a minimally-disturbing POVM is very similar to the calculations
in Sec.~\ref{Sec 5} which shows the close formal 
analogy of both methods.

%
%
\section{Conclusions}
We have presented a general scheme to implement minimally-disturbing
symmetric measurements by quantum circuits.  By applying it to the
Heisenberg-Weyl group, we obtain circuits for simultaneous
measurements of position and momentum of a particle moving on a
discrete cyclic chain.  We show that an infinite dimensional
generalization of this circuit leads to a well-defined measurement
process on a Schr\"odinger particle moving on the real line using two
probe particles.  The ``circuit'' for this continuous variable quantum
system can in principle be obtained by particle collisions with
hard-core potential.  The whole measurement process on the three
particles shows some analogies but also differences to simultaneous
measurements of the quadrature amplitudes in quantum optics using two
ancilla modes.

\vspace{0.5cm}
This work was supported by Landesstiftung Baden-W\"urttemberg
gGmbH (cooperation of the projects 
AZ 1.1322.01 and AZ 1.1422.01).

%
%
\appendix
\section*{Appendix}

\subsection*{Proof of Th.~\ref{cyc povm}}  
The $(2n \times 2)$-matrix $M$ of Eq.~(\ref{matrix m}) has the
$(\sigma_\pi \otimes \sigma,\sigma)$-symmetry that is defined by
\[
(X_n \otimes R_n) M = M R_n, 
\]
as straightforward computation shows, i.e., we have $(\sigma_\pi
\otimes \sigma)(j)=X_n^j \otimes R_n^j$ and $\sigma(j)=R_n^j$. To find
the diagonalizing operations $A$ and $B$ of Th.~\ref{th erg} we
observe that $\sigma$ is already decomposed into irreducible
representations and $B$ is therefore trivial. To decompose $\sigma_\pi
\otimes \sigma$ we diagonalize $\sigma_\pi$ by the Fourier transform
$F_n$. The eigenvalues of $(F_n X_n F^\dagger) \otimes R_n$
are in the order $1,
\omega_n,\omega_n,\omega_n^2,\omega_n^2,\dots,\omega_n^{n-1},
\omega_n^{n-1},1$. We apply the cyclic shift $X_{2n}$ to group them
into a sequence of pairs $(\omega_n^j,\omega_n^j)$ as in 
Lemma~\ref{lem int}. Therefore, we have $A=X_{2n} (F_n \otimes I_2)$.
Following Th.~\ref{th erg} we only have to find 
$W\in {\rm Int}(A(\sigma_\pi \otimes \sigma)A^\dagger, B \sigma B^\dagger 
\oplus \sigma^\prime)$ 
which is a unitary extension of 
$N:=A^\dagger M B$.
Hence, we can choose
\[
W:=
\sqrt{\frac{1}{2}}\pmatrix{1&0 & \cdots & 0 & 1 & 0  & \cdots  & 0\cr
1&0 & \cdots & 0 & -1 & 0  & \cdots  & 0\cr
0&1 & \cdots & 0 & 0 & 1  & \cdots  & 0\cr
0&1 & \cdots & 0 & 0 & -1  & \cdots  & 0\cr
\vdots & \vdots & & \vdots &\vdots &\vdots && \vdots \cr
0&0 & \cdots & 1 & 0 & 0  & \cdots  & 1\cr 
0&0 & \cdots & 1 & 0 & 0  & \cdots  & -1} 
\,,
\]
since the first two columns of this matrix coincide with $N$. This is
verified by straightforward computations, too.  One can also easily
check that $W$ can be written as $W=(I_n \otimes F_2)K^\dagger$ where
$K$ is defined in Th.~\ref{cyc povm}.

\subsection*{Proof of Th.~\ref{thm hw}} 
To define $M$ as in Eq.~(\ref{matrix m}) we have to define a
correspondence between ancilla basis states and POVM operators. Since
our ancilla system is a tensor product of two $d$-dimensional systems,
this correspondence is canonical and we obtain
\begin{equation}\label{Eq 42}
M=\sum_{j,k=0}^{d-1} \ket{j} \otimes \ket{k} \otimes
Z^k_d X_d^j \sqrt{\mu} X_d^{-j} Z_d^{-k}
\in {\mathbbm C}^{d^3 \times d}.
\end{equation}
The symmetry $(\sigma_\pi \otimes \sigma)M=M\sigma$ of $M$ is defined by
\[
\left(I_d \otimes X_d \otimes Z_d\right) M = M Z_d \hbox{\quad and \quad} 
  \left(X_d \otimes I_d \otimes X_d\right) M = M X_d.\,,
\]
as straightforward computations show.
Following Th.~\ref{th erg} we decompose the representation on
the left side into a direct sum of irreducible representations.
First of all, we diagonalize the shifts $X_d$ in the first and second tensor 
components by 
the Fourier transform. We obtain
\[
\left(I_d \otimes Z_d \otimes Z_d \right) 
  \left(F_d \otimes F_d \otimes I_d\right) M = 
  \left(F_d \otimes F_d \otimes I_d\right) M Z_d
\]
and
\[
\left(Z_d \otimes I_d \otimes X_d \right) 
  \left(F_d \otimes F_d \otimes I_d\right) M = 
  \left(F_d \otimes F_d \otimes I_d\right) M X_d.
\]
The matrices on the left side can be written as
\[
\left(I_d \otimes Z_d \otimes Z_d\right) = \bigoplus_{j=0}^{d^2-1} 
\omega_d^{j \, {\rm mod}\, d} Z_d \hbox{\quad and \quad} 
\left(Z_d \otimes I_d \otimes X_d\right) = \bigoplus_{j=0}^{d^2-1} 
\omega_d^{j \,{\rm div}\, d} X_d.
\]
Therefore, the representation
is decomposed into a direct sum of representations that are
equal to $\sigma$ up to phase factors. We now eliminate these
factors.
To simplify notation we define the block diagonal matrices
\[
X_{\rm mod} := \bigoplus_{j=0}^{d^2-1} 
X_d^{j \,{\rm mod}\, d} \quad {\rm and} 
\quad Z_{\rm div} := \bigoplus_{j=0}^{d^2-1} Z_d^{j \,{\rm div}\, d}.
\]
Using $Z_d^\dagger X_dZ_d= \omega_d^{-1} X_d$ and 
$X_dZ_dX_d^\dagger = \omega_d^{-1} Z_d$ we obtain 
\[
X_{\rm mod} 
\left(\bigoplus_{j=0}^{d^2-1} 
\omega_d^{j \, {\rm mod}\, d} Z_d\right)
X_{\rm mod}^\dagger
= \bigoplus_{j=0}^{d^2-1} Z_d
\]
and
\[
Z_{\rm div}^\dagger
\left(\bigoplus_{j=0}^{d^2-1} 
\omega_d^{j \, {\rm div}\, d} X_d\right)
Z_{\rm div}
= \bigoplus_{j=0}^{d^2-1} X_d.
\]
Using both equations we can write
\[
X_{\rm mod} Z_{\rm div}^\dagger
\left(I_d \otimes Z_d \otimes Z_d \right)
Z_{\rm div} X_{\rm mod}^\dagger
=\left(I_d \otimes I_d \otimes Z_d\right)
\]
and
\[
X_{\rm mod} Z_{\rm div}^\dagger
\left(Z_d \otimes I_d \otimes X_d \right)
Z_{\rm div} X_{\rm mod}^\dagger
=\left(I_d \otimes I_d \otimes X_d\right)\,,
\]
where we have no phase factors.  Consequently, we obtain
\[
\left(I_d \otimes I_d \otimes Z_d \right) 
X_{\rm mod} Z_{\rm div}^\dagger
\left(F_d \otimes F_d \otimes I_d\right) M = 
X_{\rm mod} Z_{\rm div}^\dagger
\left(F_d \otimes F_d \otimes I_d\right) M Z_d
\]
and
\[
\left(I_d \otimes I_d \otimes X_d \right) 
X_{\rm mod} Z_{\rm div}^\dagger
\left(F_d \otimes F_d \otimes I_d\right) M = 
X_{\rm mod} Z_{\rm div}^\dagger
\left(F_d \otimes F_d \otimes I_d\right) M X_d.
\]
We can rewrite this as 
\begin{equation}\label{Eq 41}
\left(I_d \otimes I_d \otimes \sigma \right) N = N \sigma
\end{equation}
with $N = X_{\rm mod}Z_{\rm div}^\dagger (F_d\otimes F_d\otimes I_d)M$. 
Hence, using the notation of Th.~\ref{th erg} we have 
\[
A:= X_{\rm mod}Z_{\rm div}^\dagger (F_d\otimes F_d\otimes I_d)
\quad {\rm and} \quad B=I_d
\]
since $\sigma$ is an irreducible representation.
The matrix $N$ is an element of 
the intertwining space ${\rm Int}(\oplus_{j=0}^{d^2-1} \sigma,\sigma)$. 
Following Lemma~\ref{lem int} it has
the decomposition 
\[
N=\ket{\Phi_1} \otimes I_d \in {\mathbbm C}^{d^3 \times d}
\]
with $\ket{\Phi_1} \in {\mathbbm C}^{d^2}$. Elementary but cumbersome 
computations\footnote{ 
Write $\sqrt{\mu}=\sum_{j=0}^{d-1} X_d^j \Delta_j$ with 
appropriate diagonal matrices
$\Delta_j$ and powers of the shift $X_d$.} show 
\begin{equation}\label{Eq 5}
\ket{\Phi_1}= (F_d^\dagger \otimes I_d)
\left( \sum_{q=0}^{d-1} \ket{q}\bra{q} \otimes X_d^{-q}\right)
\left( \sqrt{d} \sum_{j,k=0}^{d-1} \sqrt{\mu}_{jk} \ket{j}
\otimes \ket{k} \right)\,.
\end{equation}
We extend the representation $\sigma$ on the right side 
of Eq.~(\ref{Eq 41}) to the direct sum
of $d^2$ copies of $\sigma$. The matrix $W$  of the resulting
intertwining space has the decomposition $C \otimes I_d$ with
$C \in {\mathbbm C}^{d^2 \times d^2}$.
Therefore, we extend $\{\ket{\Phi_1}\}$ to an orthonormal basis
$\{\ket{\Phi_1}, \ket{\Phi_2} , \ldots , \ket{\Phi_{d^2}} \}$ of
${\mathbbm C}^{d^2}$.
We can define the unitary 
\[
U:=A^\dagger W (B\oplus \tilde{B})=A^\dagger
\Big((\ket{\Phi_1} \ket{\Phi_2}  
\ldots \ket{\Phi_{d^2}}) \otimes I_d\Big) 
\]
that extends $M$ with $\tilde{B}:=I_{(n-1)d}$.
Now we show how to simplify the implementation by preparing an 
appropriate ancilla state.
We have
\[
U(\ket{0} \otimes \ket{\Psi})= A^\dagger
\Big((\ket{\Phi_1} \ket{\Phi_2} 
\ldots \ket{\Phi_{d^2}}) \otimes I_d\Big)(\ket{0} \otimes \ket{\Psi})
= A^\dagger (\ket{\Phi_1} \otimes \ket{\Psi}).
\]
Hence, we can omit the implementation of $W$  
if we initialize the ancilla with $\ket{\Phi_1}$ of Eq.~(\ref{Eq 5}). 
In summary, we have to implement the unitary
\[
(F_d^\dagger \otimes F_d^\dagger \otimes I_d) Z_{\rm div}
X_{\rm mod}^\dagger (F_d^\dagger \otimes I_d \otimes I_d)
\left( \sum_{q=0}^{d-1} \ket{q}\bra{q} \otimes X^{-q}_d \otimes I_d\right)
\]
after we have initialized the ancillas with the state  vector
\begin{equation}\label{MuVektor}
\ket{\gamma}:=\sqrt{d} \sum_{j,k=0}^{d-1} \sqrt{\mu}_{jk} \ket{j} \otimes
\ket{k} \in {\mathbbm C}^{d^2}\,.
\end{equation}

As a special case consider the initial operator $\mu = \ket{\alpha}
\bra{\alpha}/d $ with $\ket{\alpha} \in {\mathbbm C}^d$ and $\langle
\alpha | \alpha \rangle = 1$.  In this case we have
$\sqrt{\mu}=\ket{\alpha}\bra{\alpha}/\sqrt{d}$.  Furthermore, we have
\[
\ket{\gamma}=\ket{\alpha}\otimes \ket{{\overline \alpha}}.
\]

%
%

\end{document}